\documentclass[twocolumn,preprintnumbers,amsmath,amssymb]{revtex4}
\usepackage{tabularx,graphicx}

\usepackage{color}
\usepackage{hyperref}
\hypersetup{
    colorlinks=true,
    linkcolor=blue,
    filecolor=blue,      
    urlcolor=blue,
}

\usepackage{color}

\usepackage{ulem}   

\begin{document}

\newcommand{\beq}{\begin{equation}}
\newcommand{\eeq}{\end{equation}}
\newcommand{\beqn}{\begin{eqnarray}}
\newcommand{\eeqn}{\end{eqnarray}}
\newcommand{\bmath}{\begin{subequations}}
\newcommand{\emath}{\end{subequations}}
\newcommand{\bra}[1]{\langle #1|}
\newcommand{\ket}[1]{|#1\rangle}


\title{On a recent explanation of the  dynamics of the Meissner   effect within the conventional theory of superconductivity}

\author{J. E. Hirsch}
\address{Department of Physics, University of California, San Diego,
La Jolla, CA 92093-0319}

  \begin{abstract} 
In Ref. \cite{hlubina}, Markos and Hlubina  argue that ``contrary to the expectations of Hirsch'' \cite{book} the conventional theory of superconductivity
  correctly describes the dynamics of the Meissner effect.  Here I point out the  flaws in their arguments that render them invalid,
  and  propose an experiment to shed further light on these issues.  
\end{abstract}
 \maketitle

  \section{introduction}
  The Meissner effect is the expulsion of magnetic field from the interior of a metal cooled into the superconducting state.
  We consider only type I superconductors in this paper. Contrary to the generally held  view \cite{tinkham}, in a series of papers \cite{meissnerpapers} beginning in 2003 \cite{lorentz} I have argued that the Meissner effect cannot occur in the absence of radial charge flow \cite{meissnerreview}. 
  The conventional theory
  of superconductivity \cite{tinkham} does not describe  radial charge flow in the transition to superconductivity, hence I have argued that it does not explain the Meissner effect.
  
  Recently, Markos and Hlubina in Ref. \cite{hlubina} have challenged my  claims that such new physics, which they characterize as ``exotic'', is required to explain the
  dynamics of the  Meissner
  effect. They  model the transition by a generalized time-dependent
  London equation, and by solving that equation together with Maxwell's equations for a variety of scenarios, both numerically and
  analytically, they describe the dynamics of the Meissner effect without any radial charge flow. Thus, they conclude that I have
  identified problems  in the conventional theory that are non-existent and that in fact  the conventional theory describes the dynamics of the Meissner effect (and the related Becker-London effect for rotating superconductors).
  
  In this paper  I point out the fatal flaws in Markos-Hlubina's arguments and discuss  an experiment to clearly expose the fallacy of their claims.

\section{Markos-Hlubina theory and its key flaw}
Markus and Hlubina start from the London equation in the form
\beq
\mu_0 {\bf j}_s=-\frac{1}{\lambda^2}{\bf A}
\eeq
with ${\bf j}_s$ the supercurrent, $\lambda$ the London penetration depth, and ${\bf A}$ the magnetic vector potential in the London gauge.
They argue that in a time-dependent situation Eq. (1) generalizes to
\beq
\mu_0{\bf j}_s({\bf r},t)=-\frac{f({\bf r},t)}{\lambda^2}  {\bf A}({\bf r},t)
\eeq
whose time derivative is
\beq
\frac{\partial {\bf j}_s({\bf r},t)}{\partial t}=\frac{1}{\mu_0\lambda^2}[f({\bf r},t) {\bf E}-\frac{\partial f({\bf r},t)}{\partial t} {\bf A}]
\eeq
where $f{\bf r},t)$ is the superconducting fraction at position ${\bf r}$ at time $t$,
with boundary conditions
\bmath
\beq
f({\bf r},0)=0
\eeq
\beq
f({\bf r},\infty)=1
\eeq
to describe the Meissner effect, with initial magnetic field
\beq
{\bf B}({\bf r},t=0)={\bf B}_0 .
 \eeq
 \emath
They solve Eqs.  (2), (3)    together with Maxwell's equations numerically and analytically for a variety of 
assumed functions $f({\bf r},t$) satisfying the boundary conditions (4) for simple geometries, and show that in all cases
the magnetic field is expelled from the interior of the system and in the final state it  is zero except within a distance of 
order $\lambda$ from the boundaries of the sample. Hence, the magnetic field was expelled, in agreement with the Meissner effect,
and its dynamics described by the conventional theory, allegedly.

Their conclusion is both mathematically correct and physically trivial. It is completely obvious that for $f({\bf r}) = 1$ the only possible solution of 
Eq. (2) together with Maxwell's equations for a simply connected body is the Meissner state with the magnetic field excluded, and that  for $f({\bf r})=0$ 
the only possible solution of 
Eq. (2) together with Maxwell's equations and Eq. (4c) is the (normal) state with uniform magnetic field in the interior and no current circulating.
$Any$ assumed $f({\bf r},t)$ that continuously evolves in time  from Eq. (4a) to Eq. (4b) with starting condition Eq. (4c) will describe
continuously the ``expulsion'' of magnetic field from the interior of the superconductor, while at the same time shedding zero light on the physics
that gives rise to the assumed $f({\bf r},t)$. Thus, their 15 pages of text, 69 equations and 14 figures teach us 
that they know how to do math, but otherwise exactly $nothing$.

\section{more detailed analysis}
Markus and Hlubina (MH hereafter) correctly state in their introduction \cite{hlubina}  that I have raised the following four questions (among others):

{\it (i) the net force acting on the electrons at the moment when the
normal metal starts turning into a superconductor vanishes, because,
at this moment, no electric field is present in the sample. Therefore the
electrons should not start moving.

(ii) even if the electrons do start moving so that they screen the
external B-field, the Faraday induction will generate an E-field which is
oriented against the electronic current, thereby stopping the electronic
motion.

(iii) the superconducting body, if it is allowed to, rotates in an
opposite direction with respect to the direction dictated by the Faraday
E-field.

(iv) when a superconductor transforms to a normal metal in presence
of a finite magnetic field, the velocity of the Cooper pairs does not decrease at the transition point. Rather, the supercurrent converts into the normal current, which stops by a dissipative process. Therefore the phase transformation process cannot happen in a dissipationless manner.}

They claim to answer these questions as follows. (In their text, references to their Eqs. (3) and (5) mean Eqs. (2) and (3) here respectively.)

In Sect. 3.2 they write

{\it ``For the shortest times $t < L/c$ the vector potential is essentially
equal to $A\sim B_0x$, and, as a result, according to (3) a finite supercurrent
flows in the plate which screens the external field, as shown in Fig. 5.
This solves Hirsch’s problem (i) from the Introduction.''}

Well no, it doesn't solve my ``problem'' (i) because the authors have not identified what is the force that makes the electrons 
suddenly acquire momentum and generate the supercurrent. The initial value of the vector potential for the uniform field $B_0$ is of course $A=B_0x$. Their Eq. (3) (Eq. (2) here) certainly says that
${\bf A}$  generates a current as soon as electrons condense from the normal into the superconducting state and $f$ in their Eq. (3) becomes non-zero. But that is simply a tautology, it explains nothing about the physics that gives rise to that current and its momentum. Nor does it explain   how that does not violate momentum conservation.

Then they write further down in Sect. 3.2

{\it ``Before proceeding, let us note that although the electric field is
positive in the right half of the plate (in agreement with the Faraday
law), the supercurrent flows here in the negative direction, see also Fig.
1. Moreover, the absolute value of $j_s$ increases as a function of time.
The explanation of this seemingly paradoxical behavior is provided by
the acceleration equation (5): the usually absent second term in this
equation dominates. Therefore, since the superconducting fraction f
grows as a function of time, the supercurrent accelerates in the negative
y direction, as required by the Meissner effect. This solves Hirsch’s
problem (ii) from the Introduction.''}

Well no, this doesn't solve my ``problem'' (ii) either.
There is no physical force proportional to the magnetic vector potential ${\bf A}$.
The only physical force in electromagnetism is the electromagnetic Lorentz force, with one term proportional
to the electric field and the other term proportional to the cross product of
velocity (or current) and magnetic field.

 Immediately thereafter  MH write:
 
 {\it ``Moreover, by virtue of (20), also the ions would move in the
negative y direction, i.e. against the electric field, if they were allowed
to. This solves Hirsch’s problem (iii). As regards the microscopic mechanism
of the momentum transfer between the superconducting electrons
and the ions, we nevertheless agree with Hirsch: the condensate forms
by annihilating pairs of normal electrons with a finite net momentum
$2p_x$ along $+y$. Thus there is a surplus of normal electrons in the -y
direction. The normal electrons relay the resulting surplus momentum
to the lattice by scattering on impurities and/or phonons.''}

Eq. (20) in their paper is
\beq
\frac{\partial \pi_{nucl}}{\partial t}=nfeE-\frac{\partial (nf)}{\partial t}p_y
\eeq
with $\pi_{nucl}$ the momentum of the ions, $ne$ the ionic charge density, and $p_y=eA$ one half of the momentum of a Cooper pair in the
supercurrent.  The first term is the action of the Faraday electric field on the ions, the second term is the negative of the momentum acquired
by electrons condensing into the superconducting state and acquiring the momentum of the supercurrent.  Eq. (5) (their Eq. (20))  is of course correct,
but it certainly does not solve my ``problem'' (iii). It is not explained how the condensing electrons acquire the momentum of the
supercurrent  leaving the normal electrons with
an imbalance of opposite momentum, and it is not explained how the momentum of the normal electrons is transfered to the body in a $reversible$ fashion, as required by thermodynamics. Scattering by impurities and/or phonons are $not$ reversible processes.

Finally in Sect. 3.4 they write

{\it ``From our Eq. (37) it follows that  $W_B=\Delta F+Q_J$  and only a minor part of $W_B$  is produced as Joule heat. Essentially the whole energy $W_B$ has to be used to (reversibly) break the Cooper pairs into normal electrons. Let us note that, in order to break a Cooper pair, also entropy has to be supplied to the system. The necessary entropy is provided by absorbing the heat $Q_{sn}$. The total energy increase of the plate caused by disappearance of the condensate therefore is $W_B+Q_{sn}=\Delta \epsilon$, as required by thermodynamics.
Although our formalism does not allow us to study the kinetics of the conversion process between the superfluid and normal electrons, we conclude that Hirsch’s problem (iv) from the Introduction is not incompatible with the London electrodynamics [31].''}

So what they call the ``kinetics of the conversion process'', in other words how the momentum of the stopping supercurrent in the S$\rightarrow$N
transition is transferred to the body as a whole without dissipation, is not allowed to be studied by their formalism. 
So my ``problem'' (iv) is certainly not solved.

They return to this point in their conclusions, where they write:

{\it In particular, we were able to identify the role of the energy $W_B$ which is released when the superconductor changes to a normal metal in a finite magnetic field. This energy is not equal to the Joule heat produced by stopping the supercurrent. Instead, $W_B$ is essentially equal to the free-energy difference between the normal and superconducting states in absence of the magnetic field. Therefore, we believe that Hirsch’s point (iv) does not apply either. It is fair to say, however, that a detailed microscopic calculation of the entropy produced when the supercurrent stops is still missing.}

It is of course true that the kinetic energy of the supercurrent that gets released when the supercurrent stops is the condensation energy that
needs to be supplied for the superconducting electrons to go normal, this was of course known \cite{scstops} before MH  were
``able to identify'' it. How the conversion occurs is the issue. At least the authors here are candid enough to acknowledge that
``a detailed microscopic calculation of the entropy produced when the supercurrent stops is still missing''. I would like to add, even a
non-detailed qualitative description of how this happens does not exist within the conventional theory.

In summary, none of the answers given by MH  to my four questions are valid answers. The authors simply sidestep the questions
by $postulating$ Eq. (2) without any physical justification, mathematically deriving   answers from it, and they acknowledge that their formalism 
doesn't allow them to provide answers.

In their Conclusions section MH  write:

{\it ``We would like to close with the following remark. Throughout this paper we have assumed that phase coherence is already established in the superconductor, so that we could apply to it the London equations. In case of the cool-first protocols, this is well justified: the processes start from perfect equilibrium states of the superconductor and switch-ing on (small) perturbations in the form of the field or rotation did not change this starting configuration appreciably.

The situation changes, however, when dealing with the field-first or rotate-first protocols: in this case, the formation of phase coherence in the condensate is part of the process. This point has been repeatedly stressed by Hirsch. So how does phase coherence appear? This is obviously a very complicated problem, related to the Kibble–Zurek mechanism [39–41]. Obviously, an approach based on the London equations cannot really address it.''}

 Indeed, that is the crux of the matter. As the authors here candidly even if belatedly acknowledge, ``an approach based on the London equations cannot really address it''. 
 
 In summary, my four questions remain unanswered. As of today, the conventional theory has not provided
 answers, the only existing proposed answers are provided by the theory of hole superconductivity \cite{meissnerreview}.
 
Similar considerations apply to MH's treatment of the Becker-London effect of rotating superconductors.
 
\section{another  point}
Markos and Hlubina state in their conclusion

{\it  
``The crucial difference between our approach and Hirsch’s analysis is rooted in the formulation of the constitutive equations of the conventional theory. Hirsch formulates them in terms of the fields ${\bf  E}$ and ${\bf B}$,  whereas we take for the London equation Eq. (3), which expresses the supercurrent  
${\bf j}_s$  in terms of the vector potential ${\bf A}$  in the London gauge. This innocent-looking difference in formulation leads in our version of the theory to the appearance of new terms in Eqs. (4), (5) in spatially and temporally varying superconductors, which are not present in Hirsch’s formulation.''}

That is not  so. One could formulate, with the same justification as MH's,  a modified  MH  theory that starts with the equivalent of their Eq. (3) (our Eq. (2)) with field rather than vector potential:
\beq
\mu_0{\nabla \times \bf j}_s({\bf r},t)=-\frac{f({\bf r},t)}{\lambda^2}  {\bf B}({\bf r},t)
\eeq
which is London's equation for time-independent $f$.
Upon taking its   time derivative it yields
\beq
\frac{\partial}{\partial t}   \nabla \times {\bf j}_s({\bf r},t) =
\frac{1}{\mu_0\lambda^2}[-f({\bf r},t) \frac{\partial {\bf B}}{\partial t}
-\frac{\partial f({\bf r},t)}{\partial t} {\bf B}] .
\eeq
While Eq. (7) is not the same as Eq. (3) in general, it is the same for the particular case where 
$f({\bf r},t)$ is assumed to be uniform, as assumed in Sect. 3.2 of MH's paper.
I would have the same objections that I have to Eq. (3), whether for a uniform case or not: the second term on the
right-hand-side of Eq. (7) is a ``force'' that does not exist in nature.

\section{Experiment to test the Markos-Hlubina theory}

                \begin{figure} [t]
 \resizebox{8.5cm}{!}{\includegraphics[width=6cm]{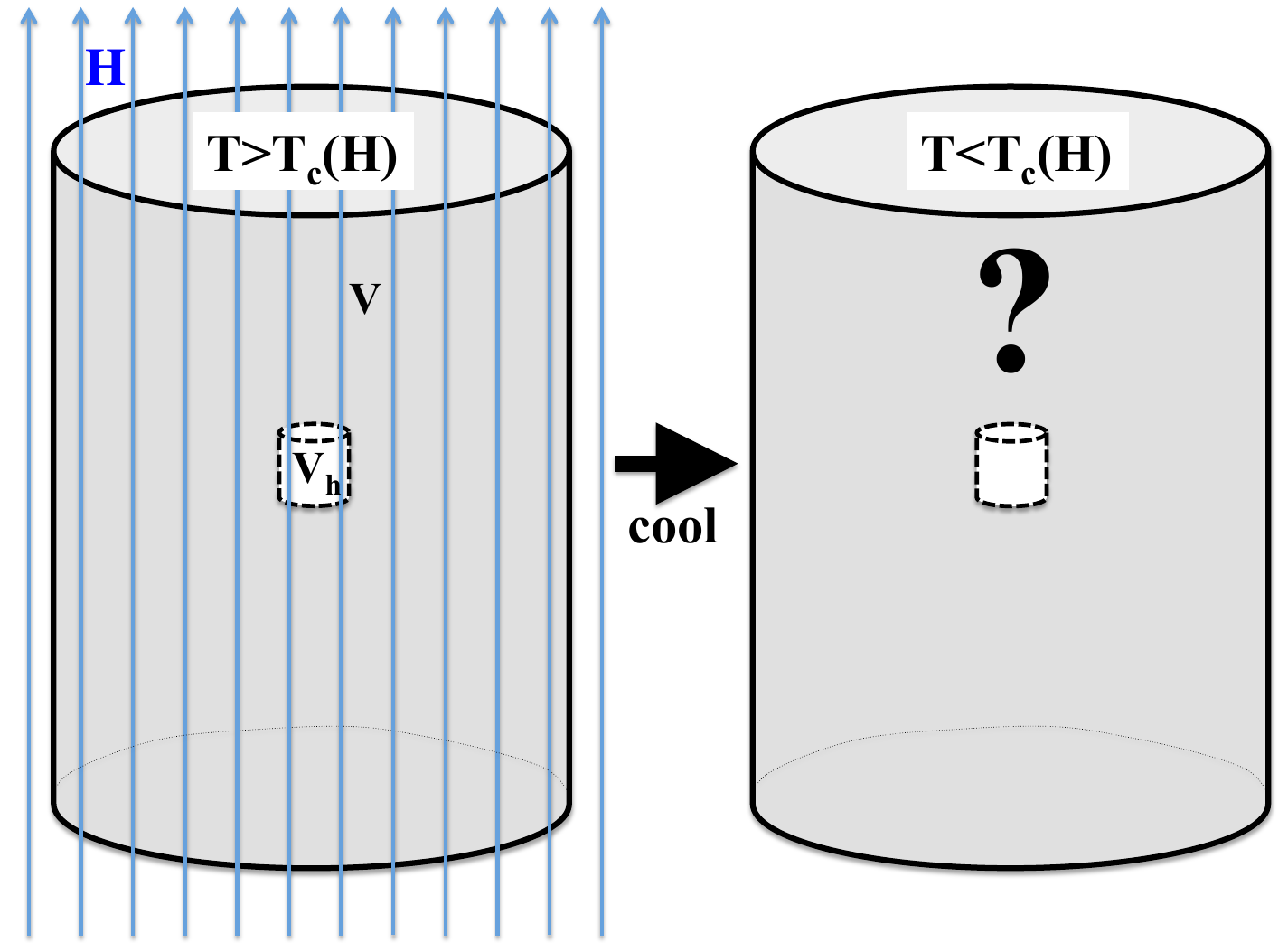}} 
 \caption {Cylindrical normal metal with a small empty  inclusion in its interior in a uniform magnetic field  that is cooled into the superconducting state.}
 \label{figure1}
 \end{figure}

 Markos and Hlubina state in their paper ``{\it we have tracked the development of electromagnetic fields for an externally prescribed condensate fraction
$f(x,t)$''}. Let us ask what they would predict in the following scenario.

Consider a superconducting cylinder of volume V with a very
  small cylindrical cavity of volume $V_h<<V$ in its interior, as shown in Fig. 1 \cite{cavity}. We ask the question: what is the final state of 
  the system when it is cooled into the superconducting state in the presence of a uniform magnetic field $\vec{B}_0=\mu_0 \vec{H}$ pointing along the
  z-axis? We assume azimuthal symmetry holds at all times. We will show 
that according to  MH's  theory the final state
is as shown in Fig. 2: current flows near the lateral surface of the cylinder, and the entire magnetic field, both from the interior
of the cylinder and its inclusion, has been expelled.

Let us start by defining the geometry of this situation. The large cylinder has radius $b$ and height $h$, extending in the $z$ direction
in the range $-h/2\le z \le h/2$. The small cylindrical cavity, centered at the center of the large cylinder, has radius $a$ and extends
in the $z$ direction
in the range $-h_h/2 \le z \le h_h/2$. We assume $h>>b$, $h_h>>a$ so we can  ignore demagnetizing effects, and $b>>a$ and $h>>h_h$.

We assume that the transition proceeds in the following stages, depicted in Fig. 3,
with the following  ``externally prescribed condensate fraction'':

{\bf Stage 1}: $0\le t\le t_1$ 
\beq f(r,z,t)=\theta(r_0(t)-r) 
\eeq
for $h_h/2<z<h/2$ and $-h/2<z<-h_h/2$, with $r_0(t)=v_1t$, with $v_1$ a constant. Then, at $t_1=a/v_1$, $r_0(t_1)=a$.

{\bf Stage 2}: $t_1\le t\le t_2$  

\beq f(r,z,t)=\theta(r_0(t)-r) 
\eeq
for $h_h/2<z<h/2$ and $-h/2<z<-h_h/2$, with $r_0(t)=v_2t$, with $v_2$ a constant, and
\beq f(r,z,t)=\theta(r_0(t)-r) - \theta(a-r)
\eeq
for $-h_h/2  \le z \le h_h/2$. Then, at $t_2=a/v_2$, $r_0(t_2)=b$. 
At this point the process ends, since in the entire region of the cylinder excluding the cavity we have $f(r,z,t_2)=1$.

To obtain quantitatively the distribution of electric fields, magnetic fields and currents for all positions and times
 it is straightforward to solve the Markos-Hlubina 
Eq. (2), together with Maxwell's equations, for the superfluid density $f(r,z,t)$ defined above. This is left as an exercise for the reader.

According to the MH theory, the superfluid distribution $f(r,z,t)$ depends on experimental conditions, and the form given by 
Eqs. (8)-(10) is certainly a plausible one, analogous the the one assumed by Markos and Hlubina in 
their Section 3.3 \cite{hlubina}. Therefore, their theory predicts that the entire magnetic field will be expelled, as shown in the right panel of Fig. 2. 
This is in agreement with the generally accepted view within BCS theory \cite{tinkham}  and Ginzburg Landau theory \cite{piers} that the system will find its way to reach its lowest
energy state, which is the state shown on  the right panel of Fig. 2.

                \begin{figure} []
 \resizebox{8.5cm}{!}{\includegraphics[width=6cm]{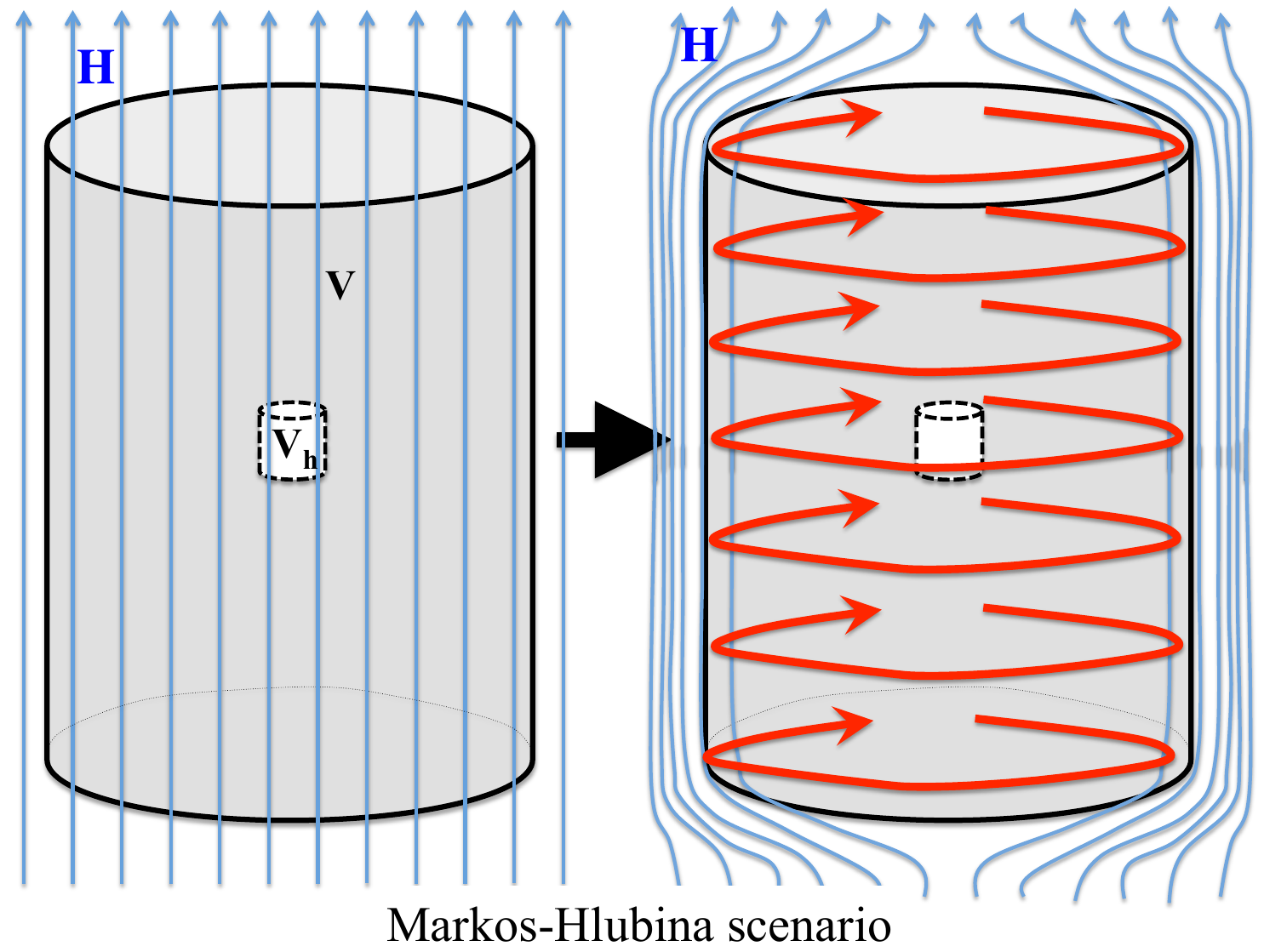}} 
 \caption {Final state when the system is cooled into the superconducting state according to the Markos-Hlubina theory. Blue lines are magnetic field lines, and red lines indicate direction of current flow.}
 \label{figure1}
 \end{figure}

                \begin{figure} []
 \resizebox{8.5cm}{!}{\includegraphics[width=6cm]{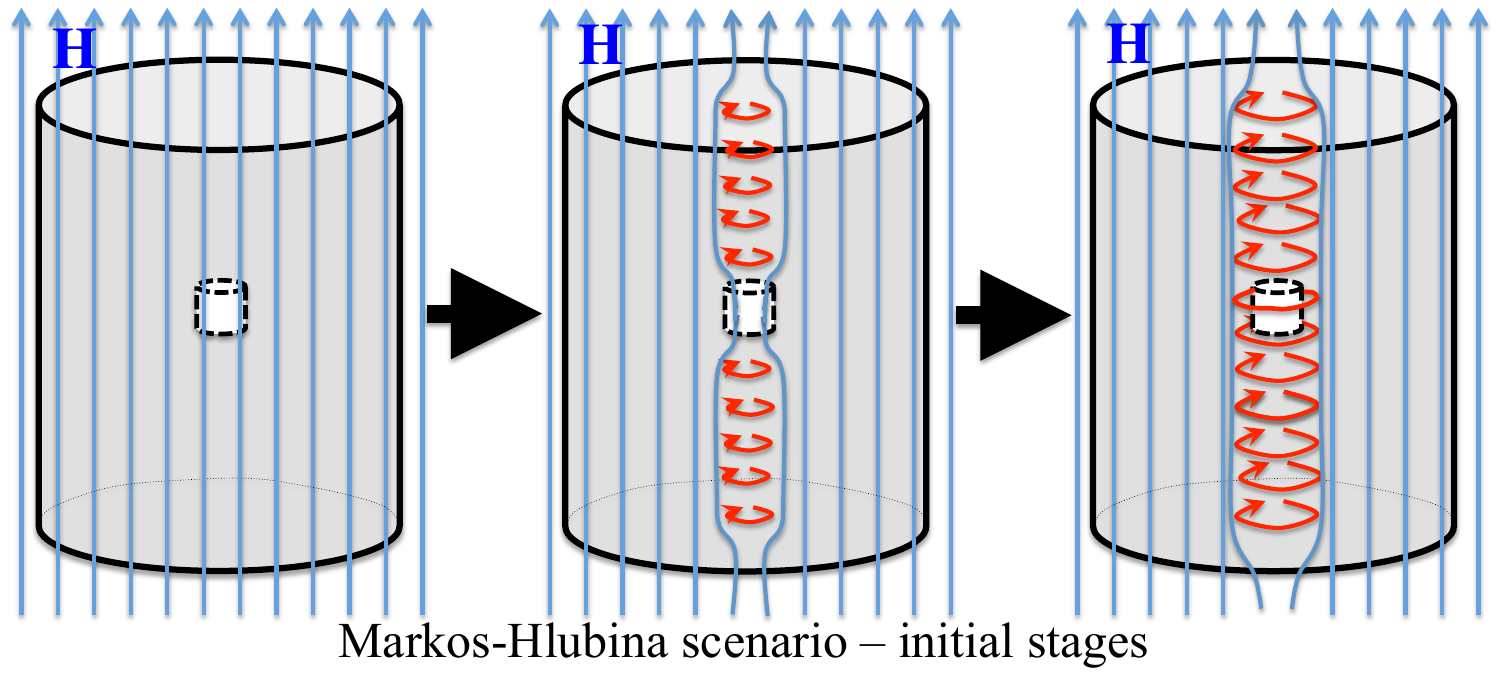}} 
 \caption {Schematic depiction of the initial stages of the transition in the Markos-Hlubina scenario. Initially,  cylindrical regions of superfluid
 nucleate above and below the inclusion, giving rise to currents that push out the magnetic field lines in that region (center panel). Once the 
 boundary $r_0(t)$ exceeds de radius  $a$ of the inclusion, superfluid also nucleates around the inclusion giving rise to currents that nullify the magnetic field in the inclusion (right panel). Upon further time evolution, the superconducting region grows uniformly across the entire height of the cylinder and magnetic field lines
 are pushed out, to reach the final state shown on the right panel of Fig. 2.}
 \label{figure1}
 \end{figure}

                \begin{figure} []
 \resizebox{8.5cm}{!}{\includegraphics[width=6cm]{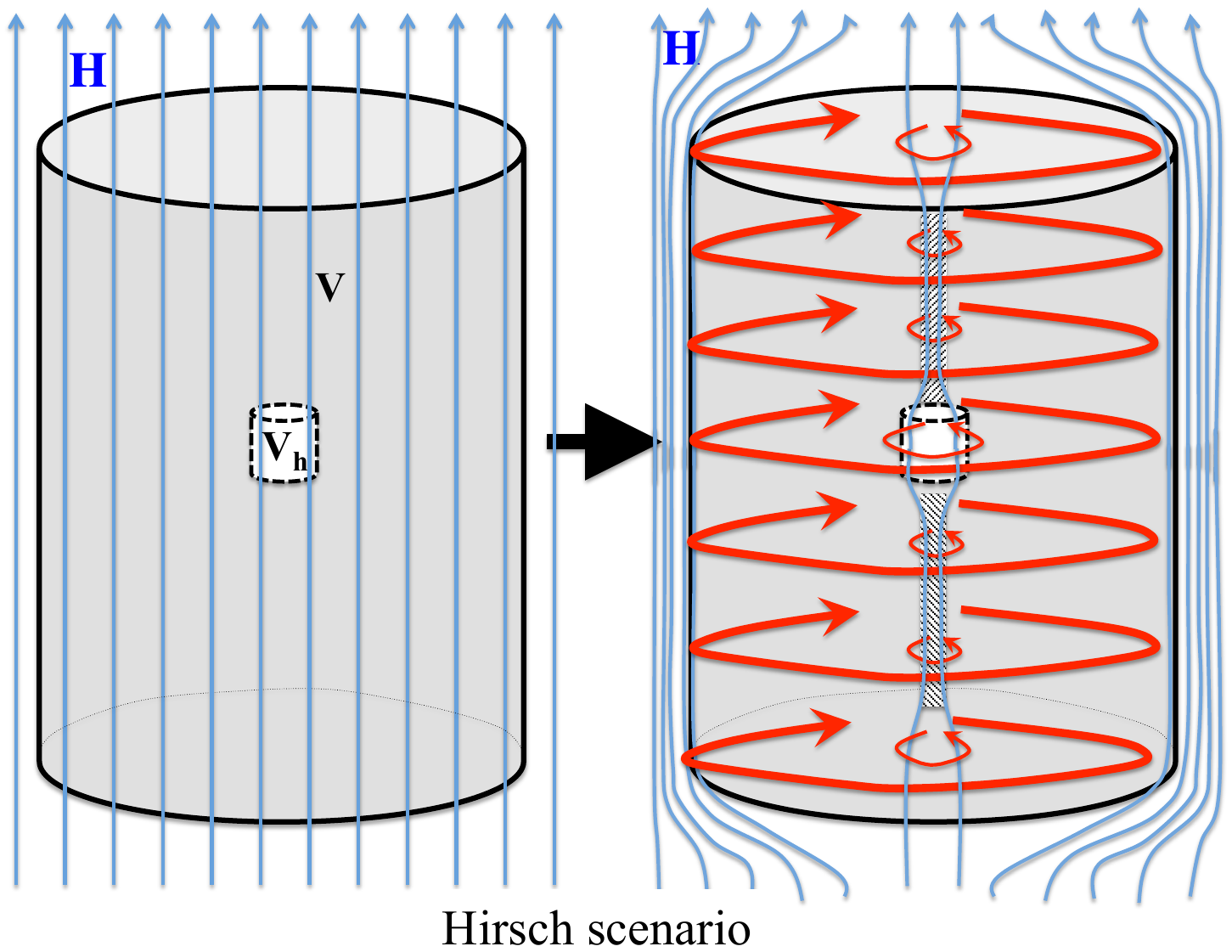}} 
 \caption { Final state when the system is cooled into the superconducting state according to the Hirsch theory. Blue lines are magnetic field lines, and red lines indicate direction of current flow. The system remains in the normal state in the regions above and below the inclusion,
 with currents flowing around it, resulting in a higher energy state.}
 \label{figure1}
 \end{figure}

\section{Alternative  view}
According to my understanding of the Meissner effect \cite{meissnerreview}, the distribution of superfluid $f(r,z,t)$ is $not$ at the experimentalist's disposal. Instead,
for superconductors to expel magnetic fields they have to also expel electric charge. More specifically, to expel the magnetic field from any 
given region of space requires outward motion of charge in that region of space \cite{cavity}. Because there is no electric charge in the
interior of the cavity, the magnetic field in the interior of the cavity  cannot be expelled. Therefore, the final state of the process has to be
what is depicted in the right panel of Fig. 4  and not what is depicted in the right panel of Fig. 2.
As shown on the right panel of Fig. 4, not all the electrons in the metallic region can transition into the superconducting state.
The magnetic field in the interior of the inclusion remains, and in order for those field lines to reach the exterior a normal region has
to exist above and below the cavity. 

The state shown on the right panel of Fig. 4 certainly has higher energy that the one shown on 
the right panel of Fig. 2, since the remaining normal regions don't benefit from the condensation energy. Nevertheless, I predict that the system will attain and remain in this metastable state: under no experimental conditions starting
with the initial state of Figs. 2 or 4 left panel would  the system be able
to reach the ground state shown on the right panel of Fig. 2, contrary to what the Markos-Hlubina theory \cite{hlubina} would predict
and more generally contrary to what the conventional theory
\cite{tinkham,piers}
predicts and contrary  to what books say that discuss this specific situation \cite{poole}.
More details are given in Ref. \cite{cavity}.
\newline
\section{conclusion}

The Markos-Hlubina theory  assumes that there are no constraints
on the possible forms of the function $f(r,z,t)$, and it does not ask what is the physics behind their Eq. (3) (Eq. (2) here). It assumes
that the vector potential uniquely determines the current in the system. It does not explain how momentum is tranferred between electrons and the
body as a whole without dissipation to account for momentum conservation. It does not ask the question of what are the forces at play that
will give rise to a given current distribution. Rather than dealing with known real  forces,  such as the Lorentz force and the
`quantum force' \cite{meissnerreview}, it assumes
a nonexistent force,   the second term in Eq. (3) here (Eq. (5) in MH's paper) that determines the dynamics, to give rise to the predictions of the conventional theory of superconductivity.

In contrast, I argue that there have to be real forces driving the dynamical behavior of the system, and that the only real
force that can give rise to azimuthal current has to derive from radial motion of electric charges, driven by quantum pressure. I furthermore argue that to interchange momentum between electrons and
the body as a whole in a reversible fashion, as required by thermodynamics,
requires electrons with negative effective mass. Since the
conventional theory does not describe  radial motion nor requires electrons with negative effective
mass, I argue that the conventional theory does not describe the dynamics of the Meissner effect \cite{meissnerreview}.

It is instructive to recall the London's view of the London equation,   as given in
F. London's 1950 book \cite{londonbook}: 
\newline
{\it ``Equations (I) and (II) in \S 3 are clearly assumptions which derive
their validity from the consequences which they imply or, rather, from
the experimental confirmation of those consequences. However, they
are not to be considered as first principles. If they are correct, it should
be possible to derive them from first principles, that is, from quantum
mechanics and from our general concepts of the structure of metals.''}
\newline
where  {``Equations (I) and (II) in \S 3''} are the London equation (Eq. (2) here, Eq. (3) in MH's paper) and Ampere-Maxwell's law, Eq. (1) in MH's paper, respectively.
Thus London, unlike MH,  clearly recognized that their equation cannot be regarded as explaining the observed Meissner effect, rather it merely describes
what is observed. MH's statement  {\it ``We demonstrate that, contrary to the expectations of Hirsch,
the conventional theory does correctly describe the dynamics of both, the Meissner and the Becker-London
effect.''}   attributes to the London equations an explanatory power that London himself explicitly denied.

The experiment discussed here offers an opportunity to shed light on these issues. If experiments show that the
state shown on the right panel of Fig. 2 is attainable, it will prove that my objections are unfounded and support the
Markos-Hlubina theory   and more generally the conventional theory.
If experiments show that it is not, it will strongly support my argument that charge expulsion is indispensable for the 
expulsion of magnetic fields in the Meissner effect. If so, the conventional theory of superconductivity will have to be discarded or amended to take
into account this fact.

 \acknowledgements
   The author is grateful to R. Hlubina  for discussions.

 \end{document}